\documentclass[lettersize, journal, 12pt]{IEEEtran}
\usepackage{amsmath,amsfonts}
\usepackage{algorithmic}
\usepackage{algorithm}
\usepackage{hyperref}
\usepackage{array}
\usepackage[caption=false,font=normalsize,labelfont=sf,textfont=sf]{subfig}
\usepackage{textcomp}
\usepackage{stfloats}
\usepackage{url}
\usepackage{verbatim}
\usepackage{graphicx}
\usepackage{cite}
\usepackage{lipsum}
\usepackage{color}
\usepackage{xcolor}
\usepackage{multirow}
\newcommand{\mr}[1]{{\color{black}{#1}}}

\begin{document}

\title{Open Experimental Measurements of Sub-6GHz Reconfigurable Intelligent Surfaces}

\author{Marco Rossanese,~\IEEEmembership{Member,~IEEE,} Placido Mursia,~\IEEEmembership{Member,~IEEE,} Andres Garcia-Saavedra,~\IEEEmembership{Senior Member,~IEEE,} Vincenzo Sciancalepore,~\IEEEmembership{Senior Member,~IEEE,} Arash Asadi,~\IEEEmembership{Senior Member,~IEEE,} and Xavier Costa-Perez,~\IEEEmembership{Senior Member,~IEEE} \\

\thanks{M. Rossanese, P. Mursia, A. Garcia-Saavedra, and V. Sciancalepore are with NEC Laboratories Europe GmbH, Heidelberg, Germany (e-mails: \{name.surname\}@neclab.eu).}
\thanks{A. Asadi is with TU Darmstadt, Darmstadt, Germany (e-mail: aasadi@wise.tu-darmstadt.de)}
\thanks{X. Costa-Pérez is with i2CAT Foundation, NEC Laboratories Europe GmbH, Heidelberg, Germany, and ICREA, Barcelona, Spain (e-mail: xavier.costa@ieee.org).}
}



\maketitle

\begin{abstract}
In this paper, we present two datasets that we make publicly available for research. The data is collected in a testbed comprised of a custom-made Reconfigurable Intelligent Surface (RIS) prototype and two regular OFDM transceivers within an anechoic chamber. First, we discuss the details of the testbed and equipment used, including insights about the design and implementation of our RIS prototype. 
We further present the methodology we employ to gather measurement samples, which consists of letting the RIS electronically steer the signal reflections from an OFDM transmitter toward a specific location. To this end, we evaluate a suitably designed configuration codebook and collect measurement samples of the received power with an OFDM receiver. Finally, we present the resulting datasets, their format, and examples of exploiting this data for research purposes. 
\end{abstract}


\newcommand{\bs}[1]{\boldsymbol{#1}}%
\newcommand{\roundP}[1]{\left(#1\right)}%
\newcommand{\squareP}[1]{\left[#1\right]}%
\newcommand{\normS}[1]{\norm{#1} ^2}%
\newcommand{\smo}[1]{\smashoperator[r]{#1}}%
\newcommand{\overbar}[1]{\mkern 1.5mu\overline{\mkern-1.5mu#1\mkern-1.5mu}\mkern 1.5mu}
\newcommand{\rmm}[1]{\mathrm{#1}}
\newcommand{\red}[1]{\textcolor{red}{#1}}
\newcommand{\st}{\mathrm{subject~to}}

\makeatletter
\let\oldabs\abs
\def\abs{\@ifstar{\oldabs}{\oldabs*}}

\let\oldnorm\norm
\def\norm{\@ifstar{\oldnorm}{\oldnorm*}}
\makeatother
\renewcommand{\a}{\mathbf{a}}
\renewcommand{\b}{\mathbf{b}}
\renewcommand{\c}{\mathbf{c}}
\renewcommand{\d}{\mathbf{d}}
\newcommand{\e}{\mathbf{e}}
\newcommand{\f}{\mathbf{f}}
\newcommand{\g}{\mathbf{g}}
\newcommand{\h}{\mathbf{h}}
\renewcommand{\i}{\mathbf{i}}
\renewcommand{\j}{\mathbf{j}}
\renewcommand{\k}{\mathbf{k}}
\newcommand{\m}{\mathbf{m}}
\newcommand{\n}{\mathbf{n}}
\renewcommand{\o}{\mathbf{o}}
\newcommand{\p}{\mathbf{p}}
\newcommand{\q}{\mathbf{q}}
\renewcommand{\r}{\mathbf{r}}
\newcommand{\s}{\mathbf{s}}
\renewcommand{\t}{\mathbf{t}}
\renewcommand{\u}{\mathbf{u}}
\renewcommand{\v}{\mathbf{v}}
\newcommand{\w}{\mathbf{w}}
\newcommand{\x}{\mathbf{x}}
\newcommand{\y}{\mathbf{y}}
\newcommand{\z}{\mathbf{z}}

\newcommand{\0}{\mathbf{0}}
\newcommand{\1}{\mathbf{1}}

\newcommand{\A}{\mathbf{A}}
\newcommand{\B}{\mathbf{B}}
\newcommand{\D}{\mathbf{D}}
\newcommand{\E}{\mathbf{E}}
\newcommand{\F}{\mathbf{F}}
\renewcommand{\H}{\mathbf{H}}
\newcommand{\I}{\mathbf{I}}
\newcommand{\J}{\mathbf{J}}
\newcommand{\K}{\mathbf{K}}
\renewcommand{\L}{\mathbf{L}}
\newcommand{\M}{\mathbf{M}}
\newcommand{\N}{\mathbf{N}}
\renewcommand{\O}{\mathbf{O}}
\renewcommand{\P}{\mathbf{P}}
\newcommand{\Q}{\mathbf{Q}}
\newcommand{\R}{\mathbf{R}}
\newcommand{\T}{\mathbf{T}}
\newcommand{\V}{\mathbf{V}}
\newcommand{\W}{\mathbf{W}}
\newcommand{\X}{\mathbf{X}}
\newcommand{\Y}{\mathbf{Y}}
\newcommand{\Z}{\mathbf{Z}}

\newcommand{\alphab}{\boldsymbol{\alpha}}
\newcommand{\betab}{\boldsymbol{\beta}}
\newcommand{\gammab}{\boldsymbol{\gamma}}
\newcommand{\deltab}{\boldsymbol{\delta}}
\newcommand{\epsilonb}{\boldsymbol{\epsilon}}
\newcommand{\varepsilonb}{\boldsymbol{\varepsilon}}
\newcommand{\zetab}{\boldsymbol{\zeta}}
\newcommand{\etab}{\boldsymbol{\eta}}
\newcommand{\thetab}{\boldsymbol{\theta}}
\newcommand{\varthetab}{\boldsymbol{\vartheta}}
\newcommand{\iotab}{\boldsymbol{\iota}}
\newcommand{\kappab}{\boldsymbol{\kappa}}
\newcommand{\lambdab}{\boldsymbol{\lambda}}
\newcommand{\mub}{\boldsymbol{\mu}}
\newcommand{\nub}{\boldsymbol{\nu}}
\newcommand{\xib}{\boldsymbol{\xi}}
\newcommand{\pib}{\boldsymbol{\pi}}
\newcommand{\varpib}{\boldsymbol{\varpi}}
\newcommand{\rhob}{\boldsymbol{\rho}}
\newcommand{\varrhob}{\boldsymbol{\varrho}}
\newcommand{\sigmab}{\boldsymbol{\sigma}}
\newcommand{\varsigmab}{\boldsymbol{\varsigma}}
\newcommand{\taub}{\boldsymbol{\tau}}
\newcommand{\upsilonb}{\boldsymbol{\upsilon}}
\newcommand{\phib}{{\boldsymbol{\phi}}}
\newcommand{\varphib}{{\boldsymbol{\varphi}}}
\newcommand{\chib}{\boldsymbol{\chi}}
\newcommand{\psib}{\boldsymbol{\psi}}
\newcommand{\omegab}{\boldsymbol{\omega}}

\newcommand{\Gammab}{\mathbf{\Gamma}}
\newcommand{\Deltab}{\mathbf{\Delta}}
\newcommand{\Thetab}{\mathbf{\Theta}}
\newcommand{\Lambdab}{\mathbf{\Lambda}}
\newcommand{\Xib}{\mathbf{\Xi}}
\newcommand{\Pib}{\mathbf{\Pi}}
\newcommand{\Sigmab}{\mathbf{\Sigma}}
\newcommand{\Upsilonb}{\boldsymbol{\Upsilon}}
\newcommand{\Phib}{\mathbf{\Phi}}
\newcommand{\Psib}{\mathbf{\Psi}}
\newcommand{\Omegab}{\mathbf{\Omega}}

\newcommand{\setA}{\mathcal{A}}
\newcommand{\setB}{\mathcal{B}}
\newcommand{\setC}{\mathcal{C}}
\newcommand{\setD}{\mathcal{D}}
\newcommand{\setE}{\mathcal{E}}
\newcommand{\setF}{\mathcal{F}}
\newcommand{\setG}{\mathcal{G}}
\newcommand{\setH}{\mathcal{H}}
\newcommand{\setI}{\mathcal{I}}
\newcommand{\setJ}{\mathcal{J}}
\newcommand{\setK}{\mathcal{K}}
\newcommand{\setL}{\mathcal{L}}
\newcommand{\setM}{\mathcal{M}}
\newcommand{\setN}{\mathcal{N}}
\newcommand{\setO}{\mathcal{O}}
\newcommand{\setP}{\mathcal{P}}
\newcommand{\setQ}{\mathcal{Q}}
\newcommand{\setR}{\mathcal{R}}
\newcommand{\setS}{\mathcal{S}}
\newcommand{\setT}{\mathcal{T}}
\newcommand{\setU}{\mathcal{U}}
\newcommand{\setV}{\mathcal{V}}
\newcommand{\setW}{\mathcal{W}}
\newcommand{\setX}{\mathcal{X}}
\newcommand{\setY}{\mathcal{Y}}
\newcommand{\setZ}{\mathcal{Z}}

\newcommand{\Real}{\mbox{$\mathbb{R}$}}
\newcommand{\Compl}{\mbox{$\mathbb{C}$}}
\newcommand{\VI}{\mathrm{VI}}

\newcommand{\argmin}{\operatornamewithlimits{argmin}}
\newcommand{\argmax}{\operatornamewithlimits{argmax}}
\newcommand{\diag}{\mathrm{diag}}
\newcommand{\Diag}{\mathrm{Diag}}
\newcommand{\diff}{\mathrm{d}}
\newcommand{\Exp}{\mathbb{E}}
\newcommand{\rmF}{\mathrm{F}}
\newcommand{\herm}{\mathrm{H}}
\renewcommand{\Im}{\mathrm{Im}}
\renewcommand{\Pr}{\mathbb{P}}
\newcommand{\rank}{\mathrm{rank}}
\renewcommand{\Re}{\mathrm{Re}}
\newcommand{\tr}{\mathrm{tr}}
\newcommand{\tran}{\mathrm{T}}
\newcommand{\Var}{\mathrm{Var}}

\section{Introduction}

Reconfigurable Intelligent Surfaces (RISs) are envisioned to become a key enabling technology for next-generation mobile systems, such as beyond-5G/6G. A RIS consists of an array of sub-wavelength unit cells that can alter the electromagnetic (EM) response of the impinging radio-frequency (RF) signals in a nearly passive way. Indeed, RISs can dynamically re-focus the received EM waves towards desired directions in space by suitably configuring the scattering properties of each unit cell. This ability unlocks new possibilities and opens up a new paradigm of the wireless environment, which has been treated as an optimization constraint in conventional systems, but can now be considered as a variable to be optimized, creating the so-called Smart Radio Environment~\cite{di2020smart}. For example, when an obstacle hinders the line-of-sight (LoS) between the transmitter and the receiver, a RIS device strategically deployed can alleviate this problem via (passive) beamforming so as to effectively create a \emph{virtual} LoS, which guarantees favorable signal propagation conditions~\cite{Mursia2020}. This can be achieved, for instance, by suitably designing the re-configurable phase shift provided by each unit cell to receive wireless signals such that the reflected signals may interfere constructively towards the desired direction and destructively elsewhere. 

\subsection{RIS requirements}
RISs are designed to be low-power, easy to manufacture, and low-cost devices. Hence, they are expected to satisfy the following requirements:
\begin{itemize}
    \item RISs should (re-)steer RF signals with minimal power loss;
    \item RISs should not use active RF components; 
    \item RISs are expected to minimize the energy required to re-configure their reflective cells;
    \item RISs should be re-configurable in real-time; and
    \item RISs are expected to be amenable to low-cost production at scale.
\end{itemize}

Several RIS prototypes have recently been presented in the literature, each of which is designed in its unique way with its associated strengths and limitations while trying to fulfill these requirements. 
However, such designs are not always completely disclosed, making their replication and the associated results hard to realize. In addition, the data collected during the characterization of these prototypes is not shared within the community. Consequently, almost the totality of works tackling RIS challenges via Machine Learning (ML) algorithms rely on simulated data. Therefore, it is evident there is a lack of prototypes to test the new possible features and scenarios and a lack of data to elaborate more accurate models for this new communication paradigm.

\subsection{Contributions} 

Motivated by these considerations, we decided to implement our RIS prototype, which fulfills the aforementioned design points, together with high granularity beam steering, and to make publicly accessible the measurements collected during our tests in an anechoic chamber, in the hope it may be useful to foster the progress of the field.\footnote{The datasets and associated documentation are available at https://github.com/marcantonio14/RIS-Power-Measurements-Dataset.} 
The design of our RIS prototype which is also briefly discussed in Sec.~\ref{sec:design} has been thoroughly explained in \cite{rossanese2022designing}. 
This paper extends our previous work by adding manifold contributions: firstly, we explain in detail how to set up an anechoic chamber for effective RIS measurements and we share our experience to ease its replication; secondly, we disclose a new dataset where we employed our unique absorption mode and we explicate how the data collected in the anechoic chamber results is structured; thirdly, we present a series of ways of how these datasets can be employed, for instance, how to model a 3D radiation pattern for the RIS from 2D measurements, how a deep neural network can help to infer missing configurations and how the absorption mode can be exploited for localization fingerprinting.

\subsection{Related Works}

This subsection gives a brief overview of existing RIS-related works, with a particular focus on datasets and some example of different prototype designs in the literature. Given the novel nature of this technology, publicly available datasets for RIS-based systems are rare today.

\mr{The only publicly accessible dataset derived from real measurements is delineated in~\cite{ds_rhur}. This dataset is constructed through the aggregation of $S_{21}$ values, denoting the transmission gain from the input to the output port, conducted at a frequency of $5$~GHz within an anechoic chamber by configuring a $16\times 16$ RIS with 1-bit phase shifter and using two distinct methodologies. It encompasses outcomes from both stationary configurations and the utilization of a rotating table, with adjustments made at intervals of $5$ degrees. Points of difference from our dataset include the absence of clearly delineated radiation patterns for the considered RIS, hindering its utility in other investigations; the provision of power values with a wider measurement interval; restriction to a single size of RIS elements; and a lack of accompanying use case illustrations for contextualization.}


\begin{table*}[]\color{black}{

\caption{Overview of Datasets related to RIS.}
\centering
\resizebox{\textwidth}{!}{%
\begin{tabular}{|c|c|c|c|c|}
\hline
\textbf{Dataset} & \textbf{Freq} & \textbf{Type} & \textbf{Description}                                & \textbf{Public} \\ \hline
- &
  5.3 GHz &
  Measurements &
  \begin{tabular}[c]{@{}c@{}}Meas. collected in anechoic chamber with different numbers of RIS elements \\ and with 3 degrees intervals. RIS radiation patterns presented.\end{tabular} &
  \checkmark \\ \hline
{~\cite{ds_rhur}} &
  5 GHz &
  Measurements &
  \begin{tabular}[c]{@{}c@{}}Meas. collected in anechoic chamber with 16×16 RIS \\ with 5 degrees intervals.\end{tabular} &
  \checkmark \\ \hline
{~\cite{elshennawy2022large}}          & Tunable       & Generic      & Generated using a custom path-loss prediction model & X               \\ \hline
{~\cite{alexandropoulos2020phase}}          & Tunable       & Generic      & Generated based on the proposed RIS channel model   & X               \\ \hline
\end{tabular}%
}

\label{tab:related_DS}
}
\end{table*}

Other existing works consist of exploiting theoretical models of RIS-aided wireless communication systems. For instance, in~\cite{elshennawy2022large}, the authors generated a dataset using a custom path-loss prediction model. Similarly, a dataset based on a proposed RIS channel model is used in~\cite{alexandropoulos2020phase}. In both cases, however, the respective datasets were not made public. \mr{The main aspects of these datasets are summarized in Table \ref{tab:related_DS} for the reader's convenience.}

Several RIS prototypes are present in the literature and they are built with different technologies and working at different frequencies~\cite{huang2022reconfigurable}. As in our case, an RF-switch at the sub-6GHz band is used in~\cite{trichopoulos2022design} with a 1-bit phase shifter. This solution is very efficient because it permits the achievement of high phase shifting granularity and low losses at limited costs; this is the main reason that drove us to choose this implementation. However, it must be said that at mmWave this technology is still immature, costly, and it can create undesired effects, such as beam squint, that must be properly addressed. 

Another phase-shifting technology is PIN-diode, which is more suited for mmWave design, as shown for the prototypes working at $28$ GHz presented in \cite{li2020novel} and \cite{dai}, while it can be used for sub-6GHz solutions as well~\cite{araghi2022reconfigurable}. Although it is a versatile solution, it presents an important limitation on the achievable phase-shifting granularity. In fact, most of the prototypes employ just 1 bit. Trying to increase this number would increase the costs together with the overall complexity since it requires the adoption of several PIN-diodes in a single unit cell. 

Furthermore, there are varactor diodes, as employed for example in \cite{pei2021ris}, which can achieve high granularity phase shifting at the cost of high power consumption. At the current stage, none of the suggested solutions can simultaneously achieve low costs, high efficiency, and low power consumption. Hence, the RIS designer needs to carefully choose the characteristics of the device based on the system constraints.

\mr{For high-frequency bands, new approaches are being investigated: for mmWave case, in \cite{cho2023mmwall}, an advanced unit cell structure working at 24.5 GHz is employed to permit the RIS to totally reflect or refract the incoming wave with the possibility to steer the output beam, while in \cite{qian2022millimirror}, a reflecting surface can be rapidly created via 3D printing to steer the wave to the desired direction. It must be noted that this solution is not reconfigurable.}

\mr{For the THz domain, instead, the research} is still in its preliminary stages, and it is far from being considered mature. This is due to the high complexity of the components and architecture, as well as the extremely small dimensions of antennas. Currently, various tuning methods are being investigated, such as electronic approaches (CMOS transistors, Schottky diodes, HEMTs, and graphene), optical approaches (photoactive semiconductor materials), phase-change materials (vanadium dioxide, chalcogenides, and liquid crystals), and microelectromechanical systems (MEMS) \cite{yang2022terahertz}.


\section{RIS design} \label{sec:design}

\begin{figure}
  \includegraphics[width=1\columnwidth]{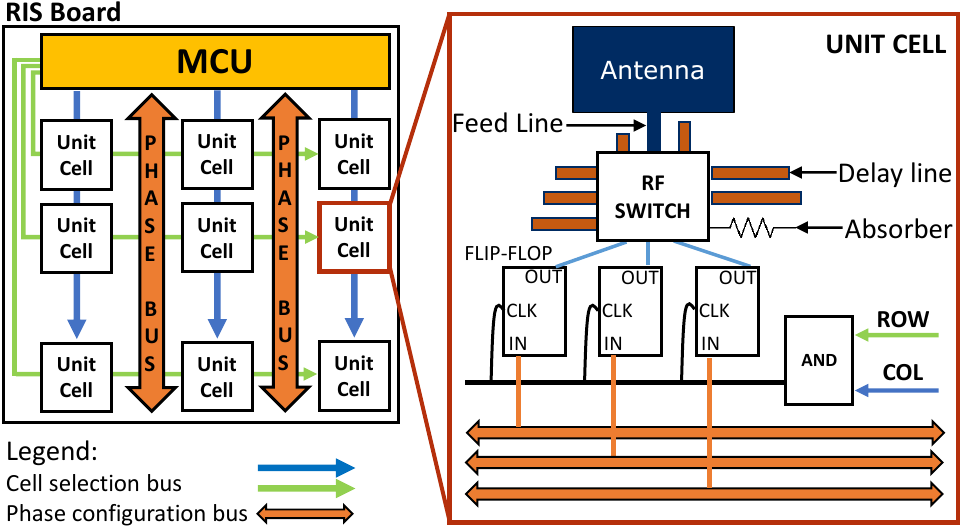}
  \caption{RIS board schematic \cite{rossanese2022designing}.}
  \label{schematic}
\end{figure}

\begin{figure*}
  \includegraphics[width=\textwidth]{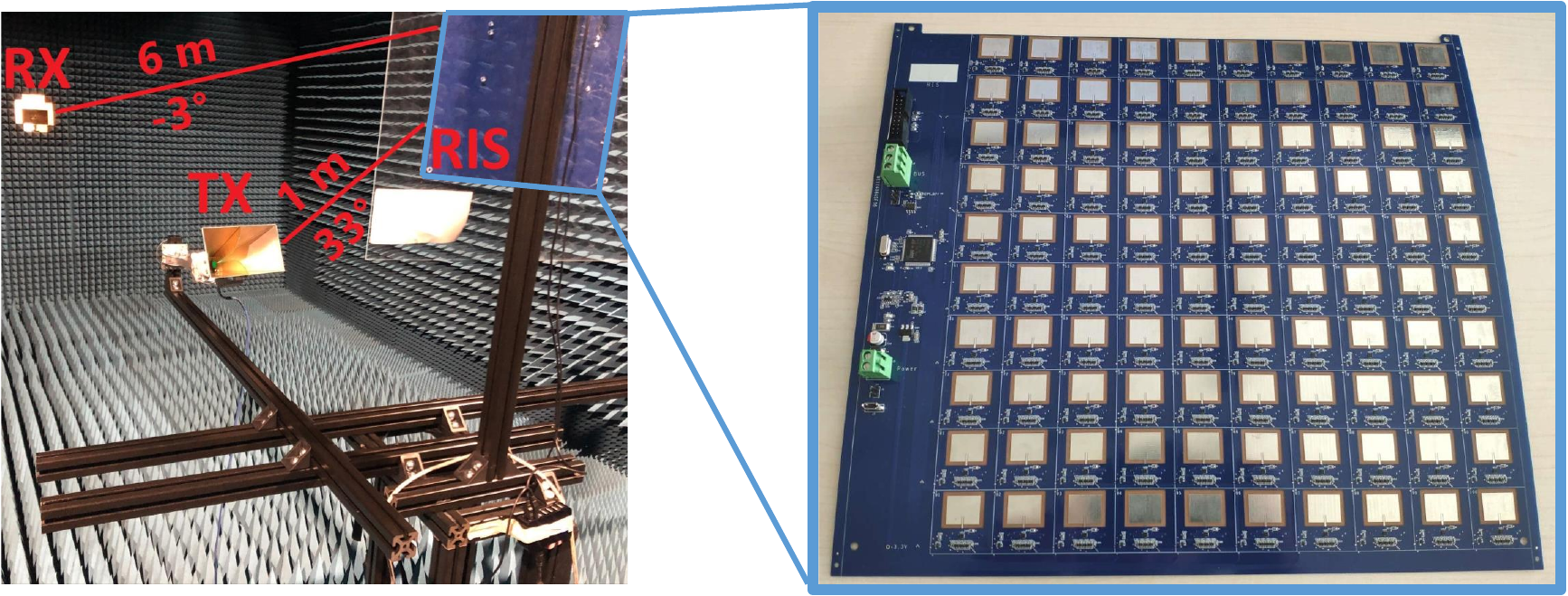}
  \caption{RIS board prototype (amplified in the right-hand side of the figure) in an anechoic chamber along with an OFDM transmitter (TX) and an OFDM receiver (RX) used to obtain the measurements provided in the dataset presented in this paper \cite{rossanese2022designing}.}
  \label{merged}
\end{figure*}

The datasets that we present in this paper have been generated by experimenting with a custom-made RIS prototype, whose design was presented in detail in~\cite{rossanese2022designing}. To provide some context, we summarize next the most important aspect of this device.

Our RIS prototype consists of $100$ patch antennas in a $10\times10$ grid of \emph{unit cells} (each with one antenna), implemented with printed circuit board (PCB) technology on a board with a thickness of $0.6$~mm. The antennas are separated by a distance equal to  $\lambda/2$, both horizontally and vertically, with $\lambda$ being the wavelength of the supported carrier signals. The operating frequency of the RIS is $5.3$~GHz, with a bandwidth equal to $100$~MHz. The final RIS prototype printout is depicted on the right-hand side of Fig.~\ref{merged}.

This prototype achieves passive beamforming via phase shifting of the impinging RF signal with $3$-bit granularity. This is implemented with suitably designed delay lines (microstrip lines in PCB): each patch antenna is connected to an RF-switch, a device that redirects an input RF signal towards one output port, while isolating the others, depending on a set configuration. The RF-switch possesses $8$ outputs, each of which is connected to an open-ended transmission line whose length determines the applied phase shift. Once the signal has traveled through the selected delay line, it is re-transmitted by the antenna. By applying the correct configuration to each RF-switch it is possible to perform passive beamforming. 

A microcontroller is in charge of configuring the RIS with the desired configuration in each unit cell. Notably, our design allows controlling each unit cell by means of a grid of configuration buses. This approach avoids the direct connection of each component to the microcontroller. Indeed, with this scheme, the total number of connections to the microcontroller is equal to the sum of rows and columns (i.e.,  $10+10$ in our case), as opposed to one connection per unit cell (i.e., $100$), thus providing a good level of scalability. A conceptual schematic of the RIS is depicted in Fig. \ref{schematic}. The average time to configure a single unit cell is $0.35$~ms and its power consumption is $62$~mW. We would like to remark that the number of unit cells has a limited impact on the power consumption since they are comprised of elementary components. On the other hand, the MCU plays a major role here because it is set to best performance mode. As part of our future work, we plan to modify the MCU to enable deep sleep mode, which will conserve power when the RIS is idle. According to the datasheet, the power consumption in sleep mode can be reduced to approximately $1 \mu W$.

The RIS is designed to be a modular device. This means that multiple RIS boards can be connected together, thus creating a larger RIS, while still respecting the $\lambda/2$ inter-element distance even between cross-board elements. 

Lastly, a peculiar characteristic of our RIS prototype is the \emph{absorption mode}, i.e., a state in which each unit cell redirects the incoming signal through one of the $8$ RF-switch outputs to a $50~\Omega$ resistor that will dissipate the incoming signal. That is, the unit cell configured with this mode will not reflect the received signal. This feature unlocks novel applications such as virtual reshaping or re-scaling of the RIS array shape in real-time.

\section{Testbed set-up and implementation}

In the following, we first present our experimental setup by describing the physical scenario, the lab equipment employed, and the measurement methodology that led to the provided datasets.

Our measurements were performed in a $5\times 8$~m anechoic chamber: a controlled environment that is isolated from external electromagnetic interference. In such a scenario, the EM waves generated by a transmitter within the chamber are absorbed by \emph{RF absorbers}, i.e., lossy material shaped to allow for incoming EM waves to penetrate with minimal reflections, which are placed on the walls, ceiling, and on the floor. Hence, the channel between the transmitter and the receiver is characterized by the direct line-of-sight (LoS) link only. Note that, for our purposes, it is essential to maintain a LoS channel, as it may be hard to distinguish between the RIS-reflected contribution and any other \emph{multipath} scattered signal components. In practice, it is not possible to block completely all other undesired paths; nevertheless, the anechoic chamber is designed to produce a reflection-free area, called \emph{quiet zone}, wherein the devices to be tested are placed.

\begin{figure}
    \centering
    \includegraphics[width=1\columnwidth]{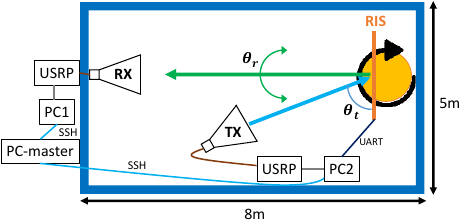}
    \caption{Anechoic chamber testbed bird's-eye view \cite{rossanese2022designing}.}
    \label{fig:bird}
\end{figure}

In our setup, we place a turntable in the quiet zone, as shown in Fig.~\ref{fig:bird}, which can be remotely controlled by a computer, dubbed as \emph{PC-master}, in a control room adjacent to the anechoic chamber. This PC has the crucial role of coordinating the transmitting antenna (TX), the receiving antenna (RX), the rotation of the table, and the RIS configuration. Indeed, it instructs two additional laptops, namely \emph{PC1} and \emph{PC2}, which are placed outside and inside the chamber, respectively, using SSH communication protocol. 
PC1 is used to collect measurements from the RX.
 Conversely,  PC2, which is connected to the RIS, receives instructions from the master PC via a Pythonn script that selects the desired RIS configuration. This communication is realized via UART communication protocol. PC2 is also connected to the TX, and it thus forwards the necessary transmission parameters received from the master PC. Lastly, the master PC is also directly connected to the turntable motor.

The RIS is installed on a pole vertically mounted onto the turntable. On the same turntable, another pole is horizontally mounted and used as a base for the TX. This setup allows us to fix the angle of arrival (AoA) of signals sent from the TX to the RIS and keep it constant for every rotation of the turntable. The TX is located at $d_{\text{RIS}-\text{TX}}=1.1$~m from the first top-left element of the RIS, with an azimuth angle of $90^\circ$, and an elevation angle of $-33^\circ$. The RX is deployed at an azimuth angle of $90^\circ$ and an elevation angle of $-3^\circ$ in front of the RIS, and it is placed at $6.3$~m away from the top-left antenna element. 

All such relative distance and angle measurements were collected with the GLM50C laser measurer and are summarized in Fig.~\ref{merged}. 
Given the RIS diagonal $D=0.43$~m and its operating frequency of $5.3$~GHz, it is hard to guarantee that the distance RIS-TX is larger than the far-field threshold, which is equal to $2\frac{D^2}{\lambda}\!=\!6.5$~m~\cite{balanis}. Nevertheless, our choice of $d_{\text{RIS}-\text{TX}}$ is larger than the \emph{reactive} near-field threshold, which is $0.62\sqrt{\frac{D^3}{\lambda}}\!=\!0.73$~m~\cite{balanis}, and thus sufficiently large for our purposes.


The TX and RX are implemented via two small double-ridged horn antennas, namely TBMA4, with a frequency range of $1 – 8$~GHz, that exhibit a gain of $13.5$~dBi and a voltage standing wave ratio (VSWR) of $\sim\!\!1$ at our operating frequency. Furthermore, its nominal impedance is $50~\Omega$.

The signal feeding the TX is generated by a dual-channel transceiver USRP model B210, offering continuous RF coverage in the range of $70$~MHz to $6$~GHz. On the RX side, another B210 USRP is used to sample and decode the incoming signal. The USRPs run the srsRAN software, i.e., an open-source SDR 4G/5G software suite from Software Radio Systems (SRS), which can process OFDM LTE-like RF signals. More specifically, the TX-side USRP is used to generate a continuous stream of OFDM QPSK-modulated symbols with $5$~MHz of bandwidth, a transmit power of $-30$~dBm per subcarrier, and numerology that meets the 3GPP LTE requirements. Whereas, the RX-side USRP measures the reference signal received power (RSRP), averaged across the signal bandwidth. In particular, the srsRAN version at the RX side was modified to conveniently dump the measured RSRP values into a file that will be later parsed and analyzed.

\begin{figure}
    \centering
    \includegraphics[width=.9\columnwidth]{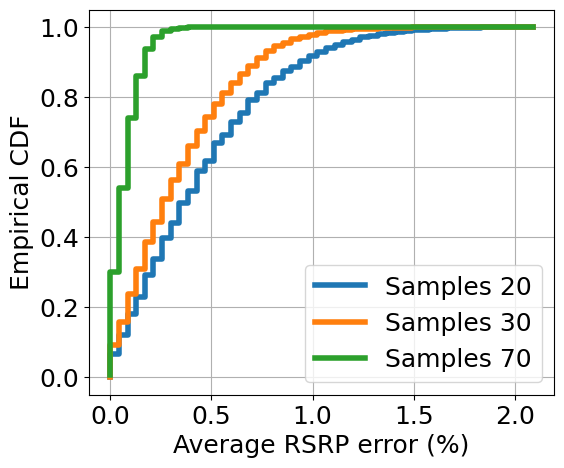}
    \caption{Empirical CDF for the RSRP averaged over different number of samples as compared to the ground truth.}
    \label{fig:error}
\end{figure}

Considering the quasi-static nature of the channel in the anechoic chamber, a small number of RSRP samples should be enough to obtain meaningful statistics. However, random fluctuations in RSRP are possible due to several factors, such as imperfect chamber isolation, intrinsic receiver noise, etc. Increasing the number of samples per measurement may help alleviate this problem, but dramatically increases the experimentation time. In order to strike a balance between speed and reliability of the measurements, we performed an initial experiment by pointing the turntable towards the RX and setting the RIS to sweep over a codebook of $3721$ predefined RIS configurations. We collected a total of $80$ RSRP samples for every RIS configuration, with a rate of $4$~ms. Given the LTE coherence time of $1$~ms and the slow-changing channel in the anechoic chamber, we selected a sampling rate that is large enough to avoid inter-symbol interference. By assuming that RSRP averaged over the $80$ samples represents the \emph{ground truth}, we then analyzed the average RSRP over a variable (and lower) number of samples, as depicted in Fig.~\ref{fig:error}. Here, we show the empirical Cumulative Distribution Function (CDF) for the difference between the average RSRP over a variable number of collected samples versus the value averaged over $80$ samples. Based on these results, it is possible to see that the average RSRP error is smaller than $1\%$ for the $90\%$ of cases when 20 samples only are taken; similarly, the error drops to $0.7\%$ when the amount of samples is 30. Therefore, we empirically chose $30$ samples, in order to keep the measured RSRP error on average below $1~\%$, and simultaneously reduce the total time of experimentation. 


We performed our tests using two predefined RIS configuration codebooks, both of which are designed by assuming a geometrical \emph{phase-shift} channel model, whereby the received signal from the RIS to a specific location in space identifyed by the azimuth and elevation angles $(\theta_r,\phi_r)$ is given as~\cite{channel_model}
\begin{align*}
    y = \h(\theta_r,\phi_r)^\herm \Phib \g(\theta_t,\phi_t) +n  \in \Compl,
\end{align*}
where $\h(\theta_r,\phi_r)\in\Compl^{N\times 1}$ is the channel from the ($N$-element) RIS to the RX, $\Phib=\mathrm{diag}(e^{j\phi_1},\ldots,e^{j\phi_N})\in\Compl^{N\times N}$ is the diagonal matrix containing the RIS phase shifts, $\g(\theta_t,\phi_t) \in \Compl^{N\times 1}$ is the channel from the TX, located at azimuth and elevation angles $(\theta_t,\phi_t)$, to the RIS, and $n$ is the noise term. Since our experiments are performed within an anechoic chamber, we adopt a purely line-of-sight model for the channel vectors as
\begin{align*}
    \h(\theta_r,\phi_r) & = \a_x(\theta_r) \otimes \a_y(\phi_r) \\
    & = [1,e^{j2\pi \delta \cos(\theta_r)},\ldots,e^{j2\pi \delta (N_x-1) \cos(\theta_r)}]\nonumber \\
    & \otimes [1,e^{j2\pi \delta \sin(\phi_r)},\ldots,e^{j2\pi \delta (N_y-1) \sin(\phi_r)}]
\end{align*}
with $\delta=0.5$ the ratio between the inter-element spacing and the signal wavelength, $N_x$ and $N_y$ the number of antennas on the rows and columns of the RIS, respectively, and $\otimes$ the Kronecker product. Note that $\g(\theta_t,\phi_t)$ is defined similarly.

In the first case, each configuration is created to point the main beam of the RIS reflection pattern toward a precise and unique direction in space. In particular, the main beam is scanned on the azimuth in the range of $[-90^{\circ}, 90^{\circ}]$ and $[-45^{\circ}, 45^{\circ}]$ on the azimuth and elevation, respectively, with a step of $3^{\circ}$ in both cases. Hence, the codebook contains a total of 1891 different configurations. 

The turntable is configured to move in the range $[-90^{\circ}, 90^{\circ}]$ with a step of $3^{\circ}$ along the azimuth plane. As stated above, the angle between the surface of the RIS and the RX along the azimuth plane is denoted by $\theta_r$. For each value of $\theta_r$, which corresponds to an equal rotation angle of the table, we let the RIS scan through all the configurations in the codebook, and we collect the RSRP power measurements at PC1, as described above. The total time for the complete characterization is $9$ hours. Given the high directionality expected in the reflections generated by the RIS (passive beamforming), we expect the RSRP to be maximum when the main beam is pointing towards the RX and very small elsewhere. However, besides verifying the directionality of our prototype, we are interested in assessing the impact of side lobes.

The aforementioned measurement procedure is replicated for two different TX locations, defined with $\theta_t$, equal to $20^{\circ}$ and $90^{\circ}$, respectively, where the latter corresponds to having the TX exactly in line with the RIS.

In the second codebook, the absorption mode feature is used to create subarrays of NxN active elements located on the top-left of the RIS, leaving all remaining antenna elements turned off. In this scenario, the table is not rotating and the receiver, transmitter, and RIS are all aligned. The main beam is scanned as in the previous case, accounting for fewer active antenna elements.

Lastly, in order to properly scale the collected RSRP measurements, we evaluated the noise floor in the anechoic chamber at a level of $-90$~dbm by collecting the RSRP when the RIS is plugged-off.
\section{Datasets and their usage}

After the required data post-processing (e.g., parsing the RSRP dump files), we consolidated the measurements we gathered in the anechoic chamber in two datasets that we made publicly available. 

In the first dataset, we used the full RIS array while rotating the turning table (dubbed as \emph{beampattern dataset}); in the second one, the table is fixed with RX, TX, and RIS aligned, while we vary the number of active antenna elements (namely, \emph{absorption mode dataset}). 

We believe these datasets can be very useful to the community. Hence, in the following, we describe them in detail and present different use cases that exemplify how this data can be exploited for research.

\subsection{Beampattern dataset}
In the former case, for each RIS configuration and rotation angle of the table, the $30$ collected samples are averaged and stored in the associated cell of a matrix: each row corresponds to the main beam direction, expressed with the couple of angles $(\theta_n, \phi_n)$, where $\theta_n$ and $\phi_n$ are the azimuth and the elevation, respectively; each column is associated with a different value of $\theta_r$, which indicates the rotation angle of the table with respect to the RX. This is illustrated in Table~\ref{tab:dataset}. Furthermore, to ease manipulation, we split this dataset into two different files for the two values of $\theta_t$, which are reflected in the file name.

Since the channel in the anechoic chamber is quasi-static, we deduced that the main factor adding noise to our RSRP measurements was related to the imperfections of the electronic components used in the RIS or the elements comprising the chamber itself. To smoothen our data, we applied a Savitzky–Golay filter, which is a common technique for smoothing data and calculations based on noisy data. Given a measured signal of $N$ points and a filter window of width $w$, i.e. a subset of N, the Savitzky–Golay filter calculates a polynomial fit of order $o$ in every window as it is moved across the signal points  \cite{Savitzky–Golay}. A good choice for our case, in which $N=61$, was using $w=7$ and $o=4$.

\begin{table*}[t]
  \centering
  \caption{Snippet of the two datasets.  On the left, the dataset for the full RIS array and rotating table (\emph{beampattern dataset}); the name of the file indicates the value of $\theta_t$. On the right, the dataset for fixed table position and different numbers of active RIS antenna elements (\emph{absorption mode dataset}).}
    \centering
	\begin{tabular}{c|ccc|}
	\cline{2-4}
	 & \multicolumn{3}{c|}{Table Rotation:  $\theta_r$} \\ \hline
	\multicolumn{1}{|c|}{Beam: ($\theta_n ; \Phi_n$)} & \multicolumn{1}{c|}{-90$^\circ$} & \multicolumn{1}{c|}{...} & 90$^\circ$ \\ \hline
	\multicolumn{1}{|c|}{(-90$^\circ$ ; -45$^\circ$)} & Power (dBm) & ... & Power (dBm) \\ \cline{1-1}
	\multicolumn{1}{|c|}{(-90$^\circ$ ; -42$^\circ$)} & Power (dBm) & ... & Power (dBm) \\ \cline{1-1}
	\multicolumn{1}{|c|}{.} & . &  & . \\
	\multicolumn{1}{|c|}{.} & . &  & .
	\end{tabular}
    \centering
	\begin{tabular}{c|cccc|}
	\cline{2-5}
	 & \multicolumn{4}{c|}{Active elements N} \\ \hline
	\multicolumn{1}{|c|}{Beam: ($\theta_n ; \Phi_n$)} & \multicolumn{1}{c|}{4} & \multicolumn{1}{c|}{16} & \multicolumn{1}{c|}{64} & 100 \\ \hline
	\multicolumn{1}{|c|}{(-90$^\circ$ ; -45$^\circ$)} & Power (dBm) & ... & ... & Power (dBm) \\ \cline{1-1}
	\multicolumn{1}{|c|}{(-90$^\circ$ ; -42$^\circ$)} & Power (dBm) & ... & ... & Power (dBm) \\ \cline{1-1}
	\multicolumn{1}{|c|}{.} & . &  &  & . \\
	\multicolumn{1}{|c|}{.} & . &  &  & .
	\end{tabular}
  \label{tab:dataset}
\end{table*}

Each row in the dataset represents the RIS reflection pattern, given a beamforming configuration, i.e., main beam orientation in space, in the range $[-90^{\circ}, 90^{\circ}]$ on the azimuth plane. Given the elevation angles between the TX and the RIS ($33^{\circ}$), and between the RIS and the RX ($-3^{\circ}$) were kept fixed during the experiments, it is possible to calculate the direction of the main beam in elevation that maximizes the RSRP: the difference between these values yields $-30^{\circ}$, which indicates that the RIS is ``virtually'' generating a main beam pointing to the ceiling. This compensates for the displacement between TX and RX on the elevation plane.

\begin{figure}
    \centering
    \includegraphics[width=1\columnwidth]{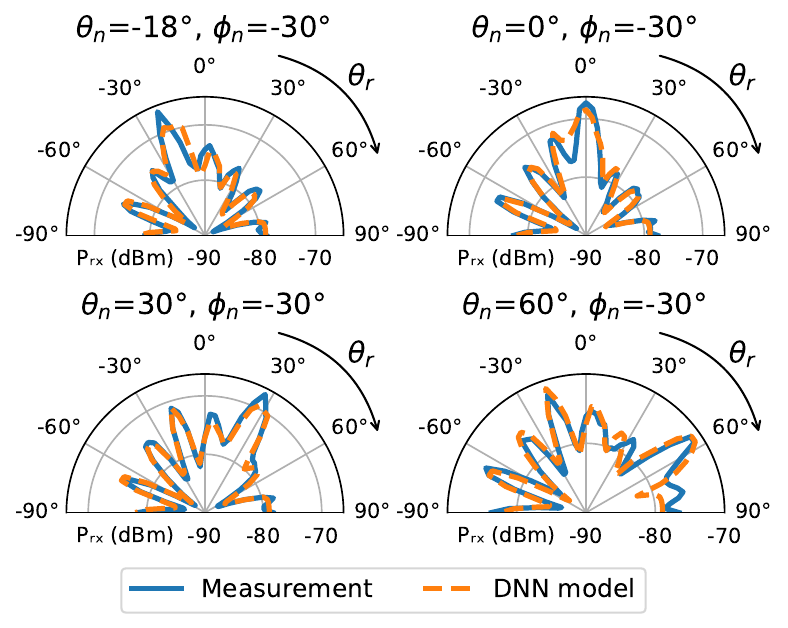}
    \caption{Power received at a fixed receiver for different values of the main beam direction in the azimuth and elevation of the RIS, represented by $\theta_n$ and $\phi_n$, respectively. The measured data is shown in blue color, while the orange lines represent the prediction via DNN.}
    \label{radpat}
\end{figure}

In Fig.~\ref{radpat}, we visualize with blue lines the rows in the dataset corresponding to azimuth angles of $\theta_n=-18^\circ$, $\theta_n=0^\circ$, $\theta_n=30^\circ$, and $\theta_n=60^\circ$ along the direction that maximizes the RSRP in the elevation, i.e., $\phi_n = -30^\circ$. The measured reflection patterns (in blue) are in accordance with what we expected for an array of patch antennas; furthermore, they demonstrate the beam-steering capabilities of our RIS prototype. By using this data, it is possible to retrieve simple, yet fundamental, parameters related to the RIS, such as the Half Power Beam Width (HPBW) or the Radar Cross-Section (RCS). 


This data is amenable to machine learning models. To demonstrate this, we trained a simple Deep Neural Network (DNN) to learn the radiation pattern of our RIS prototype for each configuration in the codebook. In particular, we generated a DNN, consisting of $3$ layers of $16$ neurons each. The chosen loss function is the mean-squared error (MSE), while Adam was used as the optimizer. We re-organized the dataset in order to feed it to the DNN by arranging the values of $\theta_n$, $\phi_n$, $\theta_r$, and the measured RSRP as columns. This new array is then split into $2$ subsets for training and validation of the DNN. The network is trained for $750$ epochs and a batch size set to $100$ rows of the dataset. 

This simple model approximates our dataset very accurately, with a normalized MSE equal to $0.01\%$ for the training set and $0.03\%$ for the validation set. 
To visualize this, Fig.~\ref{radpat} depicts with orange lines some example values in comparison to the ground truth (in blue): the DNN can accurately predict the location and magnitude of the main beams and associated side lobes, as well as the null points. This model can also be used to infer the missing configurations due to the chosen 3$^\circ$ step.

\begin{figure}
    \centering
    \includegraphics[width=0.6\columnwidth]{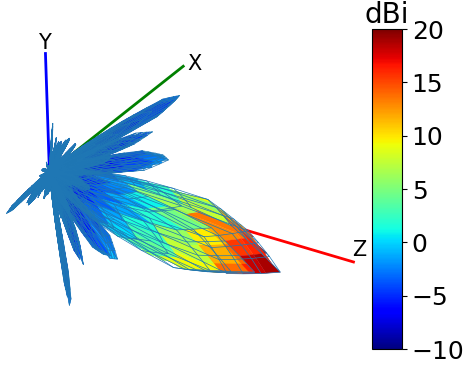}
    \caption{3D reconstruction of the RIS radiation pattern using the HPI algorithm and the provided dataset.}
    \label{plot3D}
\end{figure}

Another useful application of the dataset is the reconstruction of a 3D RIS reflection pattern. Generally, to accomplish this task, it is necessary to possess information on two orthogonal planes, but in our case, we can rely only on the azimuth plane. However, given the squared geometry of our prototype array of patch antennas, it is possible to exploit the symmetry between azimuth and elevation planes for the radiation pattern. By exploiting the interpolation algorithm called Horizontal Projection Interpolation (HPI), which is suited for the radiation shows an electrical tilt, i.e. when the main beam is not exactly perpendicular to the radiating device ~\cite{aman}. The obtained result is shown in Fig.~\ref{plot3D}.

\subsection{Absorption mode dataset}
Regarding the second dataset, we fixed the rotating table at $\theta_r=0$ and exploited the absorption mode feature of our RIS prototype, thereby reshaping the number of active antenna elements to $4$, $16$, $64$, and $100$. The main purpose of this test is to explore the relationship between the beamforming gain and the number of antenna elements. 

\begin{figure}
\centering
  \includegraphics[width=.9\columnwidth]{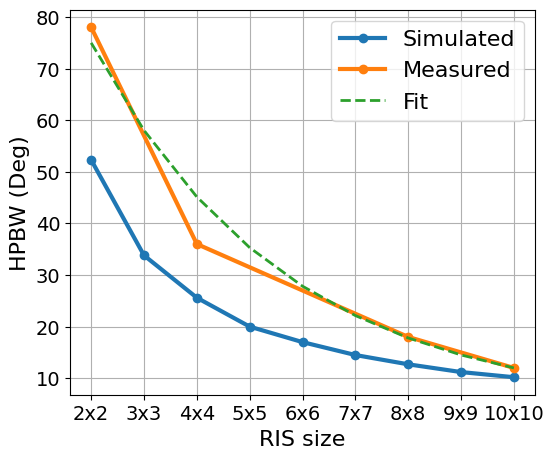}
  \caption{Half Power Beamwidth for different RIS sizes collected in the anechoic chamber (orange color), simulated via CST (blue color), and an exponential fit to the measured data (green color).}
  \label{hpbw}
\end{figure}

In this regard, we propose a model for Half Power Beamwidth (HPBW) estimation as a function of the number of (active) RIS elements. Indeed, it is well known that the theoretical computation of the HPBW for a planar array is a difficult task~\cite{balanis}. However, as shown in orange color in Fig.~\ref{hpbw}, it appears that the HPBW measured from our RIS prototype follows an exponential trend. To confirm this intuition, we used the HPBW of several RIS structures computed in CST Studio as a benchmark, which is shown with a blue color. Both the measured and synthetic data exhibit a similar trend, with a gap between the two curves, which depends on several non-idealities such as, e.g., imperfect RIS hardware, and measurement uncertainty. Moreover, in green color, we show an exponential fit of the form $a \cdot \exp(b \cdot x) + c$ to the measured data, with $x$ being the RIS size and parameters $a=70.96$, $b=0.27$, and $c=3.99$ calculated with the non-linear least squares method.

Our RIS prototype is modular and can be easily scaled to structures larger than $10\times 10$. This allows us to estimate the HPBW for larger RIS structures, which is important for localization. By reading the beampattern dataset in Table~\ref{tab:dataset}-(a) by columns, we can gain insights into the performance of RIS-based localization techniques. Each column represents a different user position from the RIS point of view. To estimate the angle of arrival/departure from/to the user, simply select the main beam direction corresponding to the maximum received power for every rotation angle of the table. The selected main beam direction can then be mapped to the corresponding RIS configuration codebook entry.

\section{Conclusion}
Reconfigurable Intelligent Surfaces (RIS) represent a promising technology that will play a major role in future mobile systems designs. In this paper, we have presented a dataset collected in an empirical setting comprised of an OFDM transmitter, an OFDM receiver, and a custom-built fully configurable sub-6GHz RIS prototype within an anechoic chamber. We have introduced the design and corresponding features of the RIS by presenting the methodology employed to gather empirical data in the system for a wide set of configurations from a high-resolution codebook, and we have provided examples to illustrate the usefulness of our dataset. 
We hope that this publicly available dataset collecting measurements in a RIS-aided wireless communication system will help in achieving a wide deployment of sub-6GHz RISs in the future.

\section*{Acknowledgments}
The research leading to these results has been supported in part by SNS JU Project 6G-DISAC (GA no. 101139130) and BeGREEN (GA no. 101097083).

\bibliographystyle{IEEEtran}
\bibliography{biblio}

\newpage
\section*{Biographies}
\vskip -2\baselineskip plus -1fil
\begin{IEEEbiographynophoto}{Marco Rossanese} received the B.Sc. and M.Sc. degrees in Telecommunication Engineering from Università degli Studi di Padova in 2017 and 2019, respectively. He is employed as a researcher at NEC Laboratories Europe GmbH in the 6G Networks team and he is enrolled as a Ph.D. student at the Technische Universität of Darmstadt. He focuses his work on Reconfigurable Intelligent Surfaces (RIS).

\end{IEEEbiographynophoto}
\vskip -2\baselineskip plus -1fil

\begin{IEEEbiographynophoto}{Placido Mursia} (S'18--M'21) received the B.Sc. and M.Sc. (with honors) degrees in Telecommunication Engineering from Politecnico of Turin in 2015 and 2018, respectively. He obtained his Ph.D from Sorbonne Université of Paris, at the Communication Systems department of EURECOM in 2021. He is currently a research scientist in the 6GN group at NEC Laboratories Europe. His research interests lie in convex optimization, signal processing and wireless communication.
\end{IEEEbiographynophoto}
\vskip -2\baselineskip plus -1fil

\begin{IEEEbiographynophoto}{Andres Garcia-Saavedra} received his PhD degree from the University Carlos III of Madrid (UC3M) in 2013. He then joined Trinity College Dublin (TCD), Ireland, as a research fellow until 2015.
Currently, he is a Principal Researcher at NEC Laboratories Europe. 
His research interests lie in the application of fundamental mathematics to real-life wireless communication systems.
\end{IEEEbiographynophoto}
\vskip -2\baselineskip plus -1fil

\begin{IEEEbiographynophoto}{Vincenzo Sciancalepore} (S’11–M’15–SM’19) received his M.Sc. degree in
Telecommunications Engineering and Telematics Engineering in 2011 and
2012, respectively, whereas in 2015, he received a double Ph.D. degree.
Currently, he is a Principal Researcher at NEC Laboratories Europe, focusing
his activity on reconfigurable intelligent surfaces. He is an Editor of the IEEE
Transactions on Wireless Communications.
\end{IEEEbiographynophoto}
\vskip -2\baselineskip plus -1fil

\begin{IEEEbiographynophoto}{Arash Asadi} is a research group leader at TU Darmstadt, leading the Wireless Communication and Sensing Lab (WISE). His research is focused on wireless communication and sensing for Beyond-5G/6G networks. He is a recipient of several awards, including Athena Young Investigator award from TU Darmstadt and outstanding PhD and master thesis awards from UC3M. Some of his papers on D2D communication have appeared in IEEE COMSOC best reading topics on D2D communication and IEEE COMSOC Tech Focus.
\end{IEEEbiographynophoto}
\vskip -2\baselineskip plus -1fil

\begin{IEEEbiographynophoto}{Xavier Costa-Perez} is Scientific Director at i2Cat, Head of 5G/6G R\&D at NEC Laboratories Europe, and a Research Professor at ICREA. His team generates research results which are regularly published at top scientific venues, produces innovations which have received several awards for successful technology transfers, participates in major European Commission R\&D collaborative projects and contributes to standardization bodies. He received both his M.Sc. and Ph.D. degrees in Telecommunications from the Polytechnic University of Catalonia (UPC). 
\end{IEEEbiographynophoto}

\end{document}